\documentclass[aps, prl, twocolumn, superscriptaddress, letterpaper]{revtex4-2}
\pdfoutput=1
\usepackage{graphicx}
\usepackage{amssymb}
\usepackage{amsmath}
\usepackage{txfonts}
\usepackage{bm}
\usepackage{color}
\usepackage[colorlinks=true, linkcolor=blue, anchorcolor=black, citecolor=blue, filecolor=black, menucolor=black, runcolor=black, urlcolor=blue]{hyperref}

\makeatletter
\renewcommand{\maketag@@@}[1]{\hbox{\m@th\normalsize\normalfont#1}}%
\makeatother

\begin{document}

\title{Observation of scale-free localized states induced by non-Hermitian defects}

\author{Xinrong Xie$^{1,4,5,6}$,Gan Liang$^2$, Fei Ma$^{1,4,5,6}$, Yulin Du$^{1,4,5,6}$, Yiwei Peng$^{1,4,5,6}$, Erping Li$^{1,4,5,6}$, Hongsheng Chen$^{1,4,5,6}$, Linhu Li$^{2,*}$, Fei Gao$^{1,4,5,6,\dagger}$ and Haoran Xue$^{3,\ddagger}$ \\
	\small $^{1}$\textit{Interdisciplinary Center for Quantum Information, State Key Laboratory of Extreme Photonics and Instrumentation, ZJU-Hangzhou Global Scientific and Technological Innovation Center, Zhejiang University, Hangzhou 310027, China} \\
	\small $^{2}$\textit{Guangdong Provincial Key Laboratory of Quantum Metrology and Sensing \& School of Physics and Astronomy, Sun Yat-Sen University (Zhuhai Campus), Zhuhai 519082, China} \\
	\small $^{3}$\textit{Department of Physics, The Chinese University of Hong Kong, Shatin, Hong Kong SAR, China} \\
	\small $^{4}$\textit{International Joint Innovation Center, The Electromagnetics Academy at Zhejiang University, Zhejiang University, Haining 314400, China} \\
	\small $^{5}$\textit{Key Lab. of Advanced Micro/Nano Electronic Devices \& Smart Systems of Zhejiang, Jinhua Institute of Zhejiang University, Zhejiang University, Jinhua 321099, China} \\
	\small $^{6}$\textit{Shaoxing Institute of Zhejiang University, Zhejiang University, Shaoxing 312000, China} \\
}

\begin{abstract}
	Wave localization is a fundamental phenomenon that appears universally in both natural materials and artificial structures and plays a crucial role in understanding the various physical properties of a system. Usually, a localized state has an exponential profile with a localization length independent of the system size. Here, we experimentally demonstrate a new class of localized states called scale-free localized states, which has an unfixed localization length scaling linearly with the system size.
	Using circuit lattices, we observe that a non-Hermitian defect added to a Hermitian lattice induces an extensive number of states with scale-free localization.  
	Furthermore, we demonstrate that, in a lattice with a parity-time-symmetric non-Hermitian defect, 
	the scale-free localization emerges because of spontaneous parity-time symmetry breaking.
	Our results uncover a new type of localized states and extend the study of defect physics to the non-Hermitian regime.
\end{abstract}

\maketitle
Wave localization is ubiquitous in nature, with examples ranging from Anderson localization in electronic materials~\cite{anderson1958} to optical solitons in nonlinear dielectrics~\cite{kivshar2003}. 	 The localized states (LSs) therein play a crucial role in understanding the physical properties of the system. For example, the formation of LSs in an Anderson insulator is responsible for the metal-insulator transition induced by random disorder. Localization can also enhance transport, as in the cases of topological materials where boundary-LSs protected by bulk topological invariant enable robust propagation~\cite{hasan2010, qi2011, ozawa2019, xue2022}. In non-Hermitian topological systems, LSs known as skin modes can strongly modify the spectrum of the system under open boundary conditions, leading to a breakdown of the bulk-boundary correspondence~\cite{yao2018, kunst2018}. These LSs also offer fascinating routes to the manipulation of wavefunctions and have resulted in many practical applications, such as soliton microcombs~\cite{kippenberg2018}, random lasing~\cite{cao1999}, light funneling~\cite{weidemann2020}, and multi-functional photonic circuits~\cite{xie2023}.

\begin{figure*}[ht]
	\centering
	\includegraphics[width=0.9\textwidth]{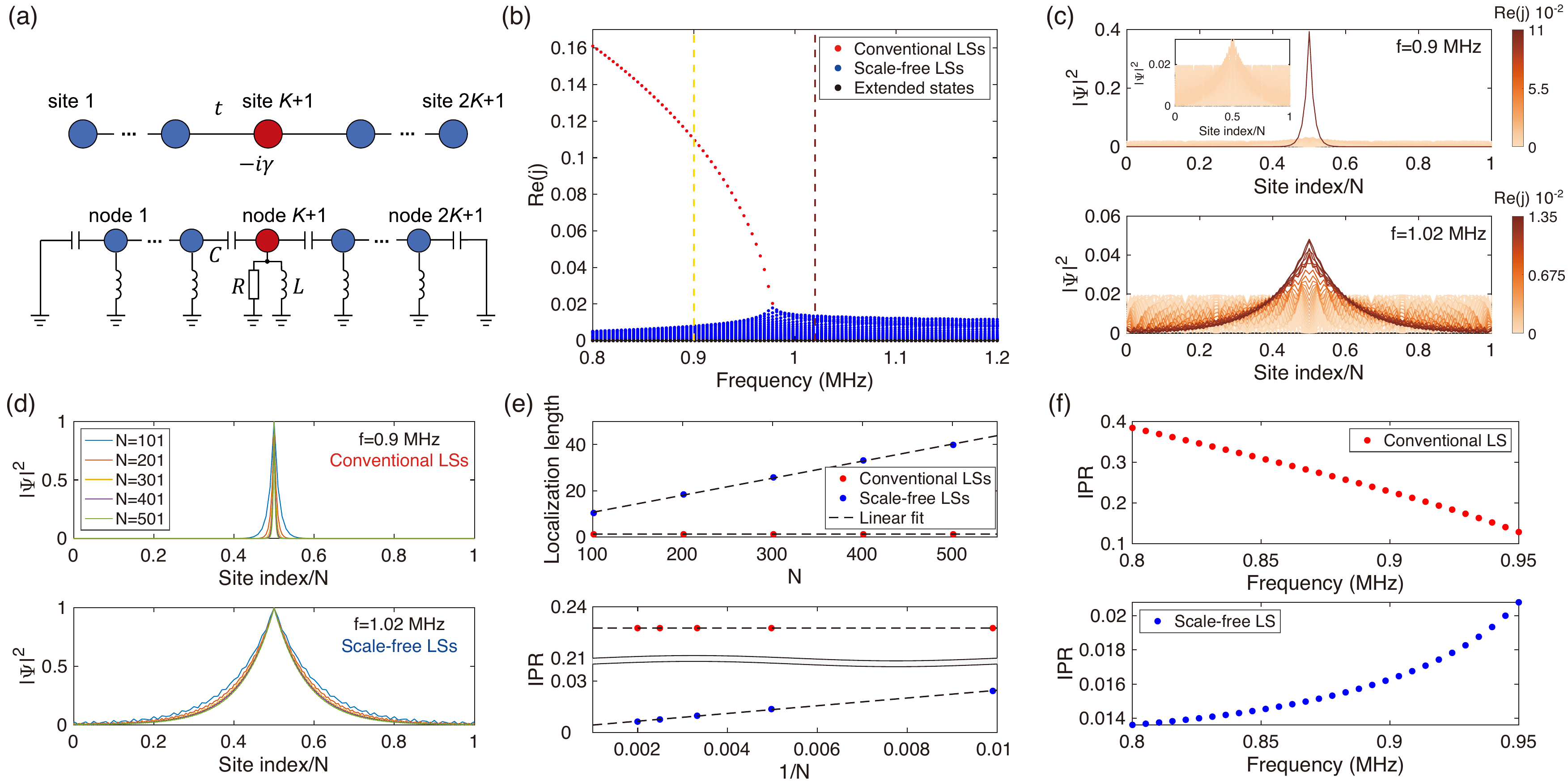}
	\caption{Scale-free localization induced by a single non-Hermitian defect. (a) Upper panel: Tight-binding model of a 1D chain with nearest--neighbor coupling $t$ and a non-Hermitian defect $-i\gamma$ at the middle of the chain (site $K+1$). Lower panel: Schematic diagram of the designed circuit realizing the tight-binding model. $C$, $R$, and $L$ denote capacitors, resistors, and inductors, respectively. (b) The real part of eigenvalues of a finite circuit chain with the fixed size $N=2K+1=101$, plotted as a function of frequency for fixed $C=22.7~\rm{nF}$, $L=1.18~\mu\rm{H}$ and $R=3.58~\Omega$. (c) Spatial distributions of all eigenstates at $0.9~\rm{MHz}$ (the upper panel) and $1.02~\rm{MHz}$ (the lower panel), as indicated by the yellow and brown dotted lines in (b). The inset in the upper panel shows the spatial distributions of eigenstates without the conventional LS. The colors denote the real part of eigenvalues. (d) Rescaled spatial distributions of conventional LSs (the upper panel) and scale-free LSs with the smallest localization lengths (the lower panel) in systems with different sizes. The horizontal axis is normalized by the system size $N$ and each eigenstate is normalized by its maximum value. (e), Plots of localization length against $N$ (the upper panel) and the IPR against $1/N$ (the lower panel) of the conventional LSs (red dots) and scale-free LSs (blue dots) shown in (d). The black dotted lines denote linear fits. (f) Defect strength dependence of localization length. The plots show the IPR against frequencies of the conventional LSs (the upper panel) and scale-free LSs with the smallest localization lengths (the lower panel) in a circuit chain with the fixed size $N=101$. Note that the increased frequency is equivalent to decreased defect strength.}
	\label{fig1}
\end{figure*}

A key characteristic of an LS is its localization length. A smaller localization length indicates a stronger localization strength. 
Conventionally, the localization length is determined by one or several system parameters, such as the defect potential for defect-induced localization,
but is independent of the system size when it is large enough to support the LS.
Recently, an exotic class of localization that does not obey this common belief has been predicted in non-Hermitian systems~\cite{li2020, li2021, yokomizo2021, molignini2023, guo2023, wang2023, li2023, fu2023, modak2023, ke2023}, 
with a localization length proportional to the system size. 
Thus, such an LS exhibits an unchanged distribution profile for systems with different sizes, dubbed as the scale-free localization~\cite{li2020, li2021}, provided the system size is normalized to unity. 
In addition to many interesting phenomena like exceptional points and the non-Hermitian skin effect that have been investigated extensively in recent years~\cite{miri2019, ding2022, lin2023},
scale-free LSs represent another unique feature of non-Hermitian systems without a Hermitian counterpart, 
and many efforts have been made in exploring their emergence in various non-Hermitian models~\cite{li2020, li2021, yokomizo2021, molignini2023, guo2023, wang2023, li2023, fu2023, modak2023, ke2023}.
However, despite these rapid theoretical advances, an experimental observation of scale-free LSs is still lacking.

In this work, we present an experimental observation of scale-free LSs in circuit lattices with non-Hermitian defects. Through measurements of the eigenstates for lattices with different sizes, the scale-free feature of the states is directly visualized. 
In addition, we experimentally uncover a series of unconventional properties of scale-free LSs that are not found in conventional LSs. Firstly, a single non-Hermitian defect can induce an extensive number of scale-free LSs, in contrast to a Hermitian defect that can only lead to one or a few LSs. We note that such a scenario also differs from the non-Hermitian skin effect where an extensive number of skin modes exist: the skin modes have size-independent localization length and are induced by a nontrivial point gap of the bulk spectrum~\cite{gong2018, okuma2020, zhang2020, borgnia2020}, while scale-free LSs do not necessarily rely on a non-Hermitian bulk and can be supported even in a Hermitian lattice with a single non-Hermitian defect. 
Secondly, the localization strength of scale-free LS is found to 
decrease monotonically with the defect strength,
in contrast to conventional LSs induced by the same defect.
Lastly, when the non-Hermitian defect is made to respect parity-time (PT) symmetry, the scale-free localization emerges because of spontaneous PT symmetry breaking (i.e., scale-free LSs exist only in the PT--broken phase). This property relates scale-free localization to the widely studied physics and applications of PT symmetry~\cite{feng2017, el2018, ozdemir2019}.

We start with a one-dimensional tight-binding model with $N=2K+1$ sites (see the upper panel of Fig.~\ref{fig1}(a)), described by the Hamiltonian
\begin{equation}
	H_1=\sum_{n=1}^{N-1}t(\hat{c}^\dagger_{n+1}\hat{c}_n+\mathrm{H.c.})-i\gamma\hat{c}^\dagger_m\hat{c}_m,\label{H1}
\end{equation}
where $\hat{c}^\dagger_n$ and $\hat{c}_n$ are the particle creation and annihilation operators, respectively, and $t$ is the nearest-neighbor coupling strength. The last term corresponds to a non-Hermitian defect at site $m=K+1$ (i.e., the middle of the chain). In the present study, we assume $\gamma$ is a positive real number, which means site $m$ is lossy. This simple model can be implemented in various passive platforms where dissipation can be engineered, such as coupled optical ring resonators~\cite{zhao2019}, laser-written waveguide arrays~\cite{weimann2017} and acoustic crystals~\cite{gao2022, gu2022}. 
Motivated by recent experimental breakthroughs in realizing quantum phases by electric circuits \cite{jia2015,victor2015,lee2018,imhof2018,wang2019,helbig2020,liu2021,dong2021,wu2022,zhang2022},
here, we utilize circuit lattices to realize this tight-binding model. 
As shown in the lower panel of Fig.~\ref{fig1}(a), the nearest-neighbor couplings are achieved through capacitors $C$, and the on-site loss is realized by a resistor $R$. 
In addition, each node is grounded by an inductor $L$.

\begin{figure*}[ht]
	\centering
	\includegraphics[width=0.9\textwidth]{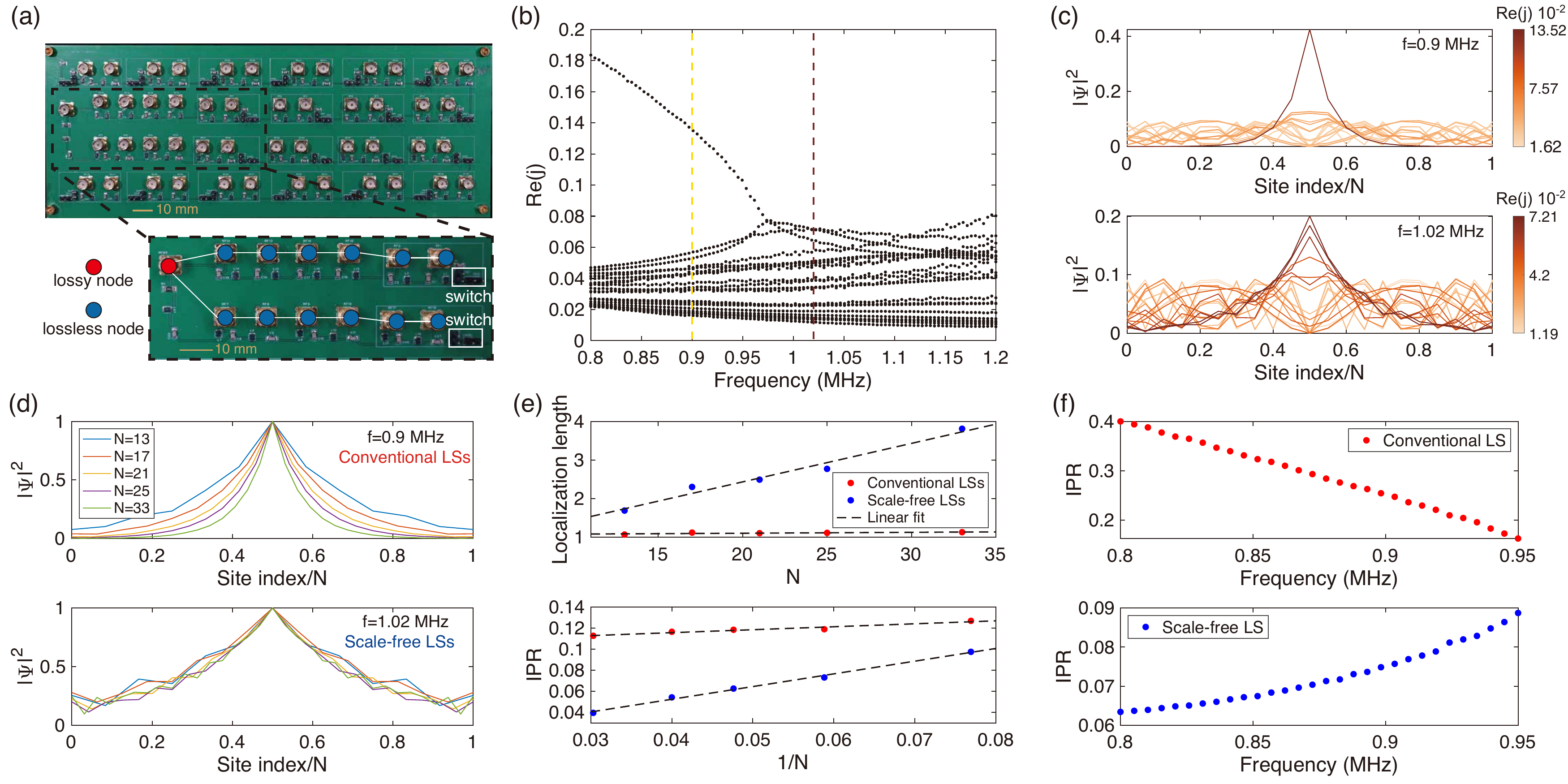}
	\caption{\textbf{Experimental observation of scale-free LSs.} (a) Photo of the fabricated circuit. The zoomed-in image shows 13 nodes where the red one denotes the lossy node and white boxes highlight switches composed of two--pin headers. (b) The real part of measured eigenvalues of a finite circuit chain with $N=21$. (c) Spatial distributions of all measured eigenstates at $0.9~\rm{MHz}$ (the upper panel) and $1.02~\rm{MHz}$ (the lower panel), as indicated by the yellow and brown dotted lines in (b). The colors denote the real part of eigenvalues. (d) Rescaled spatial distributions of measured eigenstates with the smallest localization length at 0.9~MHz (the upper panel) and 1.02~MHz (the lower panel) in systems with different system sizes. Note that the horizontal axis is normalized by the system size $N$ and each eigenstate is normalized by its maximum value. (e) Plots of localization length against $N$ (the upper panel) and the IPR against $1/N$ (the lower panel) of measured conventional LSs (red dots) and scale-free LSs (blue dots) shown in (d). The black dotted lines denote linear fits. (f) Defect strength dependence of localization length. The plots show the IPR against frequencies of the conventional LSs (the upper panel) and scale-free LSs (the lower panel) with the smallest localization lengths in a circuit chain with the fixed size $N=21$. Note that the increased frequency is equivalent to decreased defect strength.}
	\label{fig2}
\end{figure*}

According to Kirchhoff's law, the circuit model can be represented by the admittance matrix, also termed as circuit Laplacian $J(\omega)$. The circuit Laplacian describes the voltage response $V(\omega)$ to an alternating-current input $I(\omega)$ according to
\begin{equation}
	I(\omega)= (D(\omega)-E(\omega)+W(\omega))V(\omega)=J(\omega)V(\omega),
	\label{circuit-master}
\end{equation}
where $\omega$ is the angular driving frequency. $D(\omega)$ and $W(\omega)$ are diagonal matrices containing the total conductances of each node to the other nodes and the ground, respectively. $E(\omega)$ is the adjacency matrix of the conductances~\cite{lee2018, imhof2018}. For the current circuit model, its Laplacian takes the form

\begin{align}
	J_1(\omega)&=i
	\begin{bmatrix}
		0 & -{\omega}C & 0 & 0 & 0 & 0 & 0\\
		-{\omega}C & 0 & -{\omega}C & 0 & 0 & 0 & 0\\
		0 & \ddots & \ddots & \ddots & 0 & 0 & 0\\
		0 & 0 & -{\omega}C & -i\frac{1}{R} & -{\omega}C & 0 & 0\\
		0 & 0 & 0 & \ddots & \ddots & \ddots & 0\\
		0 & 0 & 0 & 0 & -{\omega}C & 0 & -{\omega}C \\
		0 & 0 & 0 & 0 & 0 & -{\omega}C & 0
	\end{bmatrix}
	\notag\\&+(2i{\omega}C+\frac{1}{i{\omega}L})M,
	\label{C_matrix}
\end{align}
where $M$ is an identity matrix of size $N$. It can be seen that $J_1(\omega)$ has a similar form to the tight-binding Hamiltonian (i.e., Eq.~\eqref{H1}) for a fixed $\omega$,
except that $J_1$ contains a global offset $(2i{\omega}C+\frac{1}{i{\omega}L})$ and an extra imaginary factor $i$. We note that
these two differences only lead to some global changes to the eigenvalues 
but have no influences on the eigenstates \cite{helbig2020}, allowing us to study the physics of $H_1$ using $J_1$. 
In the experiment, the circuit Laplacian is measured through a vector network analyzer. Therefore, we have full access to the eigenspace of $J_1(\omega)$ (see Methods). Moreover, tuning the operating frequency is equivalent to tuning the parameter $\gamma/t$ in the tight-binding model, which allows us to explore different parameter regimes without the need to change the circuit lattices. 

We further solve Eq.~\eqref{C_matrix} and present some key numerical results.
Fig.~\ref{fig1}(b) shows the real part of eigenvalues of the circuit Laplacian for a finite chain with $N=101$, plotted against frequencies for fixed $C=22.7~\rm{nF}$, $L=1.18~\mu\rm{H}$ and $R=3.58~\Omega$.
As can be seen, an extensive number of scale-free LSs (blue dots) are induced by a single non-Hermitian defect for a broad range of frequencies. When the frequency is below a critical value, $f_0=1/(4\pi{CR})\approx 0.98$ MHz (see Supplementary Data), a branch of conventional LS (red dots) bifurcates from the continuum of scale-free LSs.  
The characteristics of different types of states are clearly observed in the corresponding eigenstate profiles at two representative frequencies (0.9 MHz and 1.02 MHz) given in Fig.~\ref{fig1}(c), where we also see some extended eigenstates other than the two types of LSs. Analytically, we find that this simple model always supports $K$ extended states, $K$ scale-free states, and a single state being a scale-free/conventional LS when $f$ is larger/smaller than $1/(4\pi{CR})$ (see Supplementary Data).

A key difference between a scale-free LS and a conventional LS is the scaling behavior upon changing the system size. To investigate this property, we plot the spatial distributions of conventional LSs and the scale-free LSs (with the smallest localization length) for various system sizes in Fig.~\ref{fig1}(d). Note that the horizontal axis is normalized by the system size $N$ and each eigenstate is normalized by its maximum value. 
In this plot, scale-free LSs under different sizes retain nearly the same profile, indicating the scale-free properties (i.e., their localization length is proportional to the system size $N$). While for the conventional LSs, their profiles differ from each other.
To further quantify the scale-free property, we numerically compute the localization length and inverse participation ratio (IPR) of the states. The localization length of a state is obtained by fitting a numerical eigenstate to the profile $Ae^{-|x-x_0|/\xi}$, where $A$ is a normalization factor, $x_0$ is the center of the chain and $\xi$ is the localization length. The IPR of a state is defined as
\begin{equation}
	\text{IPR}=\frac{\sum_{x=1}^{N}|{\psi}(x)|^4}{(\sum_{x=1}^{N}|\psi(x) |^2)^2},
\end{equation}
where the index $x=1,2,\cdots,N$ is the site index. A larger IPR value indicates a more localized profile.  As depicted in Fig.~\ref{fig1}(e), the localization length of scale-free localization scales linearly with the system size, which is the hallmark of scale-free localization. This also indicates the scale-free LS becomes less localized as the system size grows and eventually transforms into an extended state ($\xi\to\infty$) when $N\to\infty$. By contrast, the conventional LS has a constant localization length with respect to the system size (Fig.~\ref{fig1}(e)). Such a difference is also seen in the IPR scaling plot, where the IPR of a scale-free LS exhibits a linear scaling to the inverse of the system size. Such a behavior is typical for an extended state. From these scaling properties, we can see that a scale-free LS is indeed a new state of matter whose localization strength lies between a conventional LS and an extended state.

In addition to the IPR and localization length scaling, another intriguing property of scale-free LSs below $f_0$ is the anomalous decrease of localization strength as the defect strength increases. Sweeping frequencies from 0.8~MHz to 0.95~MHz, we can clearly identify the relations between the localization strength (measured by the IPR) and defect strength for these two kinds of LSs. As shown in Fig.~\ref{fig1}(f), 
beyond conventional notions, the localization strength of a scale-free LS increases with increased frequency (equivalent to decreased defect strength; see Eq.~\eqref{C_matrix}). The situation for a conventional LS, as one normally expected, reverses (Fig.~\ref{fig1}(f)).
This further distinguishes scale-free localization from the conventional one.

\begin{figure*}[ht]
	\centering
	\includegraphics[width=0.9\textwidth]{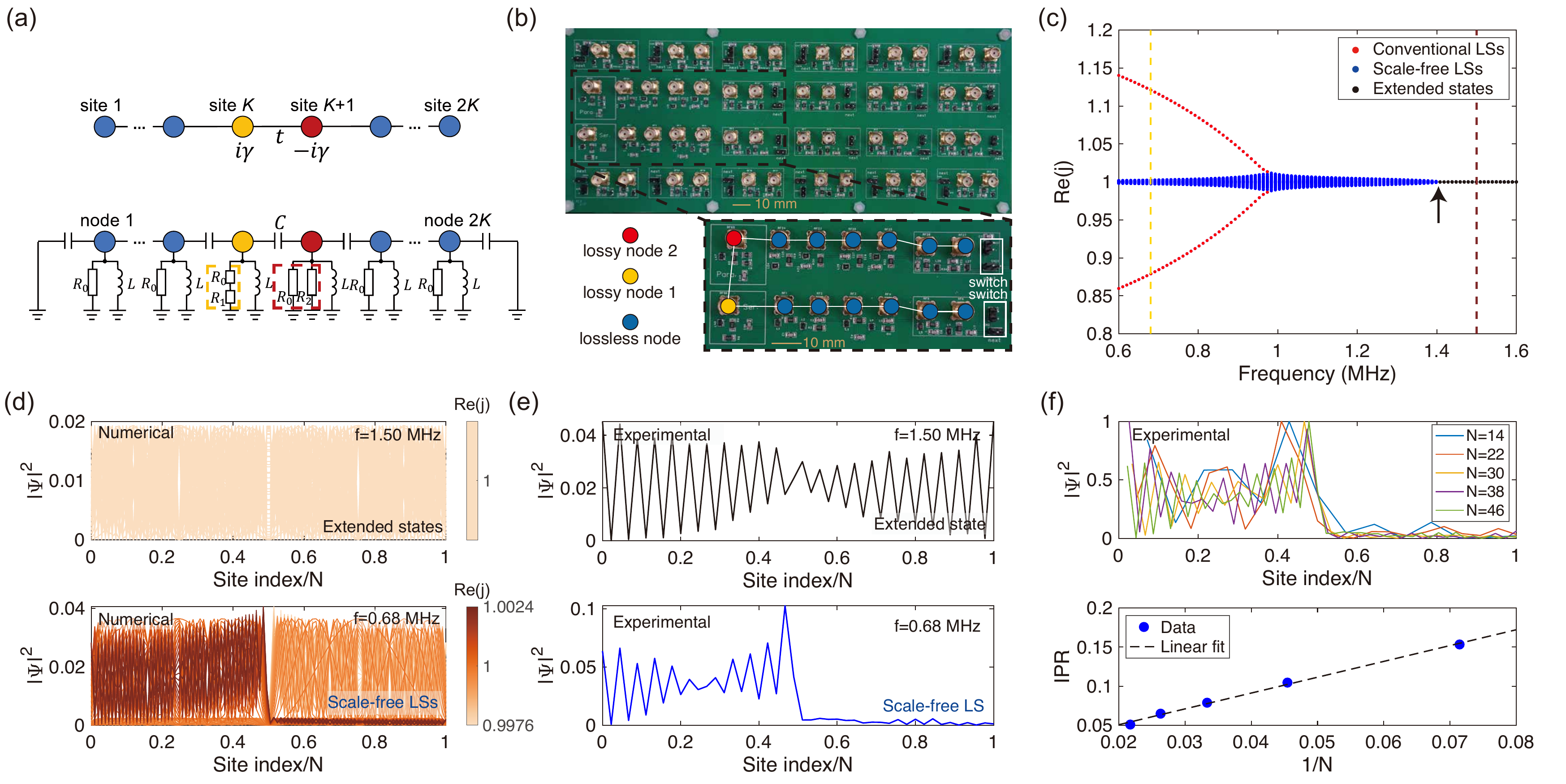}
	\caption{\textbf{Scale-free localization induced by a PT--symmetric defect.} (a) Upper panel: Tight-binding model of a 1D finite chain with nearest--neighbor coupling $t$ and two non-Hermitian defect sites ($i\gamma$ and $-i\gamma$) respecting PT symmetry. Lower panel: Schematic diagram of the designed circuit realizing the tight-binding model. The PT symmetry is preserved by setting 
		$\frac{1}{R_0}-\frac{1}{R_0+R_1}=\frac{1}{R_2}$.
		(b) Photo of the fabricated circuit. The parameters of the elements are $C=$22.7~nF, $L=$1.18~$\mu$H, $R_0=1~\Omega$, $R_1=0.25~\Omega$ and $R_2=5~\Omega$. 
		The zoomed-in image shows 14 nodes where the red and orange ones denote the two lossy nodes and the white boxes highlight switches composed of two--pin headers. (c) The real part of eigenvalues of a finite circuit chain with the fixed size $N=2K=102$, plotted as a function of frequency. The black arrow denotes the PT phase transition point. (d) Spatial distributions of all numerical eigenstates at $1.5~\rm{MHz}$ (the upper panel) and  $0.68~\rm{MHz}$ (the lower panel, with two conventional LSs excluded), as indicated by the yellow and brown dotted lines in (c). The colors denote the real part of eigenvalues. (e) Spatial distributions of two measured eigenstates ($N$=46) at $1.5~\rm{MHz}$ (the upper panel) and $0.68~\rm{MHz}$ (the lower panel), which correspond to an extended state and a scale-free LS, respectively. (f) Upper panel: Rescaled spatial distributions of the measured scale-free LSs with the largest IPR values in systems with different sizes at 0.68~MHz. Note that the horizontal axis is normalized by the system size $N$ and each eigenstate is normalized by its maximum value. Lower panel: Plot of the IPR of the measured scale-free LSs shown in the upper panel against $1/N$. The black dotted line denotes the linear fit.}
	\label{fig3}
\end{figure*}

To experimentally verify the theoretical predictions, we fabricate a size-tunable electric circuit as shown in Fig.~\ref{fig2}(a). The circuit parameters are similar to those in the theoretical model except that realistic inductors are with direct--current resistors $R_d=330~\rm{m}\Omega$, which only leads to a global offset to the eigenvalues but has no influences on the eigenstates \cite{helbig2020}. The zoomed-in image displays 13 nodes, in which the red one is the lossy node. There are several switches composed of two--pin headers (highlighted with white boxes in the zoomed-in image) in the sample, which are used to adjust the circuit size by isolating or connecting with the back-end circuit when we insert mini jumpers at different locations. With this design, we can realize several different values of $N$ on one single chip. The circuit Laplacian is experimentally obtained by measuring the $N$-port $S$-parameter of the entire work (see Methods)~\cite{zhang2022,helbig2020,liu2021}.

Figure~\ref{fig2}(b) shows the experimental eigenvalues (the real part) calculated from measured circuit Laplacian for a chain with $N=21$, which are consistent with the numerical results shown in Fig.~\ref{fig1}(b). The small deviation can be attributed to the errors of the circuit elements, the frequency--dependence of circuit elements, and the direct--current resistor of the inductor (see Methods and  Supplementary Data). The corresponding eigenstates at two specific frequencies (0.9~MHz and 1.02~MHz) are given in Fig.~\ref{fig2}(c), which reveal the coexistence of scale-free LSs and extended states at 1.02~MHz and the emergence of one extra conventional LS at 0.9~MHz.

To further confirm the nature of the observed LSs, we use the switches to adjust the size of the circuit and repeat the measurement of the circuit Laplacian for different system sizes. Fig.~\ref{fig2}(d) displays the measured eigenstates with the largest IPR values in systems with different sizes at 0.9~MHz and 1.02~MHz. 
Similar to theoretical analysis results, selected eigenstates in systems with different sizes at 0.9~MHz have different profiles (note the horizontal axis is normalized by the system size $N$), which indicates they are conventional LSs. While the profiles of the selected eigenstates at 1.02~MHz remain unchanged, showing the scale-free behavior. 
We note that there is more than one scale-free LS, as can be seen in Fig.~\ref{fig2}(c). Here we choose to focus on the one with the largest IPR value to easily trace the state as system size varies. The quantitative results of the LSs are shown in Fig.~\ref{fig2}(e). As expected, a linear scaling is found for the scale-free LS when the localization length is plotted against $N$ and the IPR is plotted against $1/N$. The former, showing the localization length is proportional to the system size, is the direct experimental evidence for the scale-free localization. Finally, we use the experimental eigenstates to uncover the anomalous relation between the localization strength and the defect strength for the scale-free localization. Similar to the theoretical analysis, we sweep the frequency from 0.8~MHz to 0.95~MHz, where conventional and scale-free LSs coexist, and select the conventional LS and the scale-free LS with the largest IPR values in each measurement. The opposite relations between the IPR and frequency are clearly observed in two kinds of LSs (see Fig.~\ref{fig2}(f)). All these experimental observations are consistent with the predictions in Fig.~\ref{fig1}.

So far, we have shown that scale-free localization can be induced simply by a single non-Hermitian defect. In fact, scale-free localization takes place in various non-Hermitian settings and can interact with other non-Hermitian phenomena~\cite{li2020, li2021, yokomizo2021, molignini2023, guo2023, wang2023, li2023, fu2023, modak2023, ke2023}. Here, we demonstrate the emergence of scale-free localization from a PT-symmetric non-Hermitian defect and its deep connection with the PT phase transition. Consider a finite tight-binding lattice that contains $N=2K$ sites and two defects with gain $i\gamma$ and loss $-i\gamma$ at sites $m_1=K$ and $m_2=K+1$, respectively (see the upper panel of Fig.~\ref{fig3}(a)). The corresponding Hamiltonian reads:
\begin{equation}
	H_2=\sum_{n=1}^{N-1}t(\hat{c}^\dagger_{n+1}\hat{c}_n+\mathrm{H.c.})+i\gamma\hat{c}^\dagger_{m_1}\hat{c}_{m_1}-i\gamma\hat{c}^\dagger_{m_2}\hat{c}_{m_2}.
\end{equation}
It is easy to check that $[H_2,PT]=0$ with $P$ a $2K\times2K$ matrix with anti-diagonal elements being $1$
and $T$  the complex conjugation. 
To avoid using active elements, we apply a background loss in the circuit design to make the system totally passive. This procedure, which is commonly adopted in studying non-Hermitian PT symmetry~\cite{guo2009, weimann2017}, will not alter the physics but only add a uniform shift to the imaginary part of the eigenvalues. The circuit design is illustrated in the lower panel of Fig.~\ref{fig3}(a), where the unequal resistors $R_1$ and $R_2$ in the middle of the circuit are introduced to realize the PT-symmetric defect, and resistors $R_0$ connected to all nodes are introduced as the global loss, which satisfies $\frac{1}{R_0}-\frac{1}{R_0+R_1}=\frac{1}{R_2}$. 
Besides, capacitors $C$ and inductors $L$ are introduced for realizing the nearest couplings and linking each node to the ground, respectively. 
According to Eq.~\eqref{circuit-master}, we can get the Laplacian of this circuit as
\begin{equation}
	J_2(\omega)=J_{2,\text{eff}}(\omega)+\frac{1}{R_0}M,
\end{equation}
where $M$ is an identity matrix of size $N$, and the circuit Laplacian apart from global loss is written as

\begin{small}
	\begin{align}
		J_{2,\text{eff}}(\omega)&=i
		\begin{bmatrix}
			0 & -{\omega}C & 0 & 0 & 0 & 0 & 0 & 0\\
			-{\omega}C & 0 & -{\omega}C & 0 & 0 & 0 & 0 & 0\\
			0 & \ddots & \ddots & \ddots & 0 & 0 & 0 & 0\\
			0 & 0 & -{\omega}C & i\frac{1}{R_2} & -{\omega}C & 0 & 0 & 0\\
			0 & 0 & 0 & -{\omega}C & -i\frac{1}{R_2} & -{\omega}C & 0 & 0\\
			0 & 0 & 0 & 0 & \ddots & \ddots & \ddots & 0\\
			0 & 0 & 0 & 0 & 0 & -{\omega}C & 0 & -{\omega}C \\
			0 & 0 & 0 & 0 & 0 & 0 & -{\omega}C & 0
		\end{bmatrix}	
		\notag\\&+(2i{\omega}C+\frac{1}{i{\omega}L})M
	\end{align}
\end{small}

The PT symmetry is evident through the relation $[-iJ_{2,\text{eff}}, PT]=0$.
The electric circuit we fabricate is shown in Fig.~\ref{fig3}(b). The zoomed-in image displays 14 nodes, in which the red and orange ones denote the two lossy nodes. Similar to the sample shown in Fig.~\ref{fig2}(a), switches following the same schemes are adopted to adjust the size of the measured circuit.

Through analytical calculation, we identify two phase transitions of this model that alter the spatial distribution of eigenstates, as detailed in Supplemental Information. Namely, all eigenstates are extended in the PT-unbroken phase with $1/(R_2\omega C)<1$, and scale-free LSs emerge only in the PT-broken phase with $1/(R_2\omega C)>1$. Further decreasing the frequency $\omega=2\pi f$, two conventional defect LSs emerge when $1/(R_2\omega C)>\sqrt{2}$.
These two transitions can be clearly seen in Fig. \ref{fig3}(c), which
shows our calculated eigenvalues (the real part) of a finite circuit chain with $N=46$, plotted as a function of frequency for fixed $C=22.7~\rm{nF}$, $L=1.18~\mu\rm{H}$,  $R_0=1~\Omega$, $R_1=0.25~\Omega$ and $R_2=5~\Omega$.	
A PT phase transition point around $f_0=1/(2\pi R_2 C)\approx1.4~\rm{MHz}$ is clearly observed (indicated by a black arrow). When $f>f_0$, all eigenvalues are purely imaginary, corresponding to the PT-unbroken phase. Decreasing $f$ to be lower than $f_0$, the system undergoes a phase transition into a PT-broken phase and the eigenvalues become complex. Picking two specific frequencies in different phases (indicated by the vertical dashed lines in Fig.~\ref{fig3}(c)),
we find that eigenstates are all extended in the PT-unbroken phase, while in the PT-broken phase, there emerge scale-free LSs (Fig.~\ref{fig3}(d)).
Note that two conventional LSs,
corresponding to the two bifurcated branches of eigenvalues with larger $|{\rm Re}(j)|$ when $f<1/(2\sqrt{2}\pi R_2 C)\approx 0.99$ MHz in Fig. \ref{fig3}(c),
are excluded in Fig.~\ref{fig3}(d) for a clear visualization of other states. 
Fig.~\ref{fig3}(e) displays two representative measured states in PT-symmetric and broken phases, respectively, which are consistent with the numerical results. To clearly visualize the scale-free localization, we again pick up the experimental eigenstates with the largest IPR values for systems with different sizes. As shown in Fig.~\ref{fig3}(f), the highly overlapped profiles and linear IPR scaling in experimental eigenstates prove the scale-free localization behavior. Besides, these scale-free LSs only have strong support on half of the chain, consistent with the fact that they lie in the PT-broken phase. 

In summary, we have experimentally observed scale-free LSs in circuit lattices with two kinds of non-Hermitian defects, i.e., a single defect and two defects respecting PT symmetry. The linear relationship between the localization length of scale-free LSs and the system size is verified experimentally. In addition, we reveal a series of novel properties associated with scale-free LSs, including the ability to induce an extensive number of scale-free LSs by a single defect, the anomalous relation between the localization strength and defect strength, and the locking of scale-free localization and the PT-broken phase, all of which are not found in conventional LSs. Our results highlight the unique features of scale-free LSs and provide a new route to the study of non-Hermitian physics. In the future, it would be interesting to investigate scale-free localization in higher dimensions and its interplay with other effects, such as nonlinearity, disorder, and the non-Hermitian skin effect, which are all available in the circuit platform. The current scheme that uses non-Hermitian defects to induce scale-free localization can also be applied to photonic, acoustic, and mechanical systems, where the scale-free LSs might find applications in lasing and energy harvesting.

$\ $

\noindent $^*$ \href{mailto:lilh56@mail.sysu.edu.cn}{lilh56@mail.sysu.edu.cn},\\ $^\dagger$\href{mailto:gaofeizju@zju.edu.cn}{gaofeizju@zju.edu.cn},\\ $^\ddagger$\href{mailto:haoranxue@cuhk.edu.hk}{haoranxue@cuhk.edu.hk}

\end{document}